\documentclass[pra,preprint,showpacs]{revtex4}
\usepackage{amsmath}
\usepackage{amsfonts}
\usepackage{amssymb}
\usepackage{amsthm} 
\usepackage{newlfont}
\usepackage{epsfig}
\usepackage{amscd}
\usepackage{graphicx}
\usepackage{footnote}
\usepackage{lipsum}
\usepackage{color}
\usepackage{xcolor}
\usepackage{graphicx}
\usepackage{multirow}
\usepackage{amscd}

\newcommand{\beq}{\begin{equation}}
\newcommand{\eeq}{\end{equation}}
\newcommand{\ba}{\begin{array}}
\newcommand{\ea}{\end{array}}
\newcommand{\bea}{\begin{eqnarray}}
\newcommand{\eea}{\end{eqnarray}}
\begin{document}

\begin{center}
{\large \sc \bf {Solving systems of linear algebraic equations via unitary transformations on quantum processor of IBM Quantum Experience. }
}

\vskip 15pt

{\large 
S.I.Doronin, E.B.Fel'dman and A.I.Zenchuk
}

{\it Corresponding author:} A.I.Zenchuk, zenchuk@itp.ac.ru

\vskip 8pt

{\it 
Institute of Problems of Chemical Physics RAS,
Chernogolovka, Moscow reg., 142432, Russia}

\end{center}

\begin{abstract}

We propose a protocol for solving systems of linear algebraic equations via quantum mechanical methods  using the  minimal number of qubits.  
We show that   $(M+1)$-qubit system is enough  to solve a system of $M$ equations for one of the  variables leaving other variables unknown provided that the matrix of a linear system satisfies certain conditions. In this case, the vector of input data (the rhs of a linear system) is encoded into the initial state of the quantum system. This protocol is realized on the 5-qubit superconducting quantum processor of IBM Quantum Experience 
for particular linear  systems of  three equations. 
We also show that the solution of a linear algebraic system can be obtained as the result of a natural evolution of an inhomogeneous spin-1/2 chain in an inhomogeneous external magnetic field with the input data encoded into the initial state of this chain. For instance, using such evolution in a 4-spin chain we solve a system of three equations. 
\end{abstract}

\maketitle

\section{Introduction}
\label{Section:Introduction}

The creation of quantum counterparts of classical algorithms solving various  algebraic problems, and their programming on IBM quantum computers is an important directions in development of quantum information processing. In our paper, we refer to  a problem of 
solving a linear system of algebraic equations $A{\mathbf x}= {\mathbf b}$ via the quntum-mechanical approach. 
A well known  algorithm of this kind was proposed by A.W.Harrow, A.Hassidim  and S.Lloyd (HHL algorithm) \cite{HHL}. It solves a linear system reducing the state of  input data $|b\rangle$  
(the quantum state encoding the vector ${\mathbf b}$) to  the  state proportional to $A^{-1} |b\rangle$. Essentially, this algorithm presents a specific method for inverting  the matrix $A$ of the algebraic system  using an extended quantum system and well established quantum protocols, such as Hamiltonian simulation \cite{BGCS,Ch} and phase estimation \cite{CEMM,LP} based on the quantum Fourier transform \cite {NCh,GN}. In addition, a special algorithm for preparing $|b\rangle$ in the basis of eigenvectors of $A$ is required \cite{A,SLRV}. Some applications of HHL-algorithm can be found in \cite{CJS,BWPRWL}.

The HHL algorithm consists of several steps. (i) The initial data representation (the vector ${\mathbf b}$) in the basis of eigenvectors of $A$; {(ii) exponentiation of the Hermitian operator $A$ using the Trotter formula to get the unitary operator $e^{i A t}$ with the time-parameter $t$;} (iii) applying the phase estimation to compute the (approximate) eigenvalues of $A$; (iv) rotation of an ancillary qubit  over the angle defined by the  eigenvalues of $A$; (v) inverse  phase estimation. The number of qubits involved into this algorithms is mainly defined by the desired number of decimals kept in the eigenvalues of $A$. This algorithm was realized in the optical system \cite{CWSCGZLLP} and in superconducting quantum processor Ref.\cite{ZSCetal} for  a particular linear system with $A=\left(\begin{array}{cc}
1.5&0.5\cr
0.5&1.5\end{array}\right)$
having simple eigenvalues $1$ and $2$. These eigenvalues can be encoded into the two-qubit register, therefore the HHL algorithm is implemented into the 
four-qubit computer in both cases quoted above. A modification of HHL algorithm aimed on the reducing the number of qubits involved into calculations 
was worked out in \cite{BKRLDAW}. This version is applicable if one needs only one bit of eigenvalues of $A$  be calculated by the phase estimation. In this case, the system of two equations can be solved using a three-qubit quantum computer.

{
The advantage of the HHL algorithm is in its inversion of the matrix $A$ by a quantum 
mechanical method. 
But, at present, its application to the full extend  meets some difficulties associated with the imperfection of quantum gates. For instance, the Trotter formular 
requires repeating application of certain unitary transformations. This involves a large number of gates, which reduce the accuracy. 
Thus, the implementation of the Trotter formula to exponentiate the XX-Hamiltonian  on the processor of the IBM Quantum Experience [17] shows  that  the deviation of the experimental result from the theoretically predicted one is large even for small  
Trotter numbers. The practical realization of HHL-algorithm presented  in Refs. [12-14] was proposed for the matrix $A$ of a very special form. 
These reasons motivate considering an algorithm which requires less number of quantum operations with the price of including the classical calculations.}

Our protocol differs from HHL algorithm. To solve a system of linear equations, we use the 
unitary transformation encoding the columns of the inverse matrix $A^{-1}$. Although this 
step requires {classical inversion of the matrix $A$}, this protocol has an 
advantage that {both} the number of qubits {and the number of gates} required for its implementation   do not depend on 
the accuracy of calculation and are defined only by the dimensionality of the considered 
linear system. Namely, the number of  qubits  is no more then twice as large as the number 
of equations and it can be reduced even more. 
{The accuracy of the obtained result  is defined by the accuracy of constructing the unitary matrix associated with the matrix $A^{-1}$.} If  the unitary transformation is found, it can 
be used to solve a class of linear equations having the same matrix $A$ and different 
right hand sides ${\mathbf b}$. {This kind of problems appears in many areas of 
mathematical and computational physics including algorithms for solving the systems of 
linear differential equations. }

We note that the unitary transformations at the receiver side were used in Ref.\cite{ZPLA2018} for structural restoring the elements of the density matrix transferred through the spin chain from the sender to the receiver and   to perform simple operations with the density matrix elements  \cite{ZArXiv2019}\footnote{In particular, it was shown that the unitary transformation  can be used to solve  
a system of two linear algebraic equations. However, that unitary transformation was not optimized for  solving linear systems 
and  includes 42 parameters. In our paper, we show that  the two-parameter unitary transformation can solve a system of two equations. 
}. Now we optimize such transformations for solving a particular algebraic problem. 

To minimize the quantum system needed for solving the system of $M$ linear equations, we split the protocol into $M$ steps, each one solving the system for a particular variable $x_k$ via the particular unitary transformation $U^{(k)}$, in which the $k$th row of the inverse matrix $A^{-1}$ is encoded. To find all $x_k$, $k=1,\dots,M$, we need $M$ unitary transformations 
$U^{(i)}$, $i=1,\dots,M$. In this case, it is enough to take a quantum system of  only $M+1$ qubits.  

We emphasize that using a system of $2M$ qubits we can implement a more complex unitary transformation completely solving 
the linear system for all $M$ variables $x_i$, $i=1,\dots,M$. This procedure is not considered here.

Having  a formally constructed unitary transformation $U^{(k)}$, we still have to provide a method for its realization. We show that the required unitary operator can be represented as a superposition of CNOTs (two-qubit operators) and one-spin rotations. However, involving a set of CNOTs decreases the accuracy of calculations \cite{ZRPL}. We study the  realization of our protocol on the basis of  
the 5-qubit superconducting quantum processor of IBM Quantum Experience considering systems of three equations with real matrices $A$ and columns ${\mathbf b}$.
Although the accuracy of the directly obtained result is rather poor, we introduce  a correction function which allows to compensate this disadvantage. Subtracting  this function from the measured  results we obtain the  accuracy   $\lesssim 25\%$ for $x_i^2\gtrsim 0.2$, $i=1,\dots, M$.  

Another aspect considered in this paper is the realization of the 
 unitary transformation needed for solving a given algebraic system as a natural evolution of an inhomogeneous 
spin chain governed by the XX Hamiltonian  in the inhomogeneous external magnetic field. In this case, similar to the previous one, the column ${\mathbf b}$ must be  encoded into the initial state of the particular spin chain 
 and the parameters of the Hamiltonian must be adjusted to find one of the unknowns $x_k$ in the  linear system with the given $A$. An example of a four-spin chain solving a system of three equations with the real matrix $A$ and column ${\mathbf b}$ is presented.
 {The advantage of the natural evolution is that it does not require  implementation of quantum gates using a special environment.}

The paper is organized as follows.
In Sec.\ref{Section:U} we discuss the general structure of the unitary transformation solving a system of linear equations,  find the constraint on the matrix $A$ and define a minimal  number of qubits in a quantum processor required for solving a system of $M$ equations. 
The representation of the above unitary transformation in terms of CNOTs and one-qubit rotations is described in Sec.\ref{Section:IBM} using examples of the linear systems of two and three equations. 
The implementation of our protocol on the superconducting quantum processor of IBM Q Experience is given in the same section. 
The method for solving linear systems through the natural evolution of the inhomogeneous chain under the nearest-neighbor 
XX Hamiltonian in the inhomogeneous external magnetic field is presented in Sec.\ref{Section:field}, where  an example of a system of three equations is considered. 
General conclusions are given in Sec.\ref{Section:conclusions}.

\section{Solving linear systems of algebraic equations via unitary transformation of quantum system}
\label{Section:U}
\subsection{Linear system of algebraic equations}
The inhomogeneous system of $M$ linear algebraic equations for $M$ unknowns can be written in the following form
\begin{eqnarray}\label{lin}
A{\mathbf x} ={\mathbf b},
\end{eqnarray}
where $A$ is a square $M\times M$ matrix, $ {\mathbf x}$ and  $ {\mathbf b}$ are $M$-dimensional columns of, respectively, unknowns and constants:
\begin{eqnarray}
{\mathbf x}=(x_1\;\dots x_M)^T, \;\; {\mathbf b}=(b_1\;\dots b_M)^T,
\end{eqnarray}
 the superscript $T$ means transpose. This system has the unique solution for any ${\mathbf b}$ if $\det A\neq 0$: ${\mathbf x}=A^{-1} {\mathbf b}$.

In eq.(\ref{lin}), $A$ is a fixed operator, which will be given a quantum-mechanical representation in terms of a unitary transformation. The vector 
${\mathbf b}$ is a vector of  input data, its elements must be encoded into the initial state of a quantum system as probability amplitudes. Finally,  ${\mathbf x}$ is the vector of output data. Its elements appear as probability amplitudes in a final state of the quantum system and can be  extracted through the proper measurements.

\subsection{Encoding {$A^{-1}$} into unitary transformation}
\label{Section:encoding}
The matrix $A$ of the linear system (\ref{lin}) is not unitary in general. However, if this matrix  satisfies certain conditions, then its inverse  can be encoded into the unitary transformation. In this case,  applying  such unitary transformation to a quantum system, whose pure state contains the entries of ${\mathbf b}$ as  probability amplitudes, we transform this state to the state  where the probability amplitudes equal unknowns  $x_i$, $i=1,\dots,K$. Let us obtain those conditions for $A$. 
 
First, we write the general form of a unitary $2M\times 2M$ block matrix (here $M$ is the dimensionality of $A$)
\begin{eqnarray}\label{U0}
U=\left(
\begin{array}{cc}
U^{(11)}&U^{(12)}\cr
U^{(21)}& U^{(22)}
\end{array}
\right),
\end{eqnarray}
where $U^{(ij)}$, $i,j=1,2$, are $M\times M$  
matrix blocks (non-unitary in general). 
The matrix $U$ must be such that the equation   
\begin{eqnarray}
\left(
\begin{array}{cc}
U^{(11)}&U^{(12)}\cr
U^{(21)}& U^{(22)}
\end{array}
\right) \left(
\begin{array}{c}
\mathbf b\cr
0_M
\end{array}
\right)= \left(
\begin{array}{c}
\mathbf x\cr
U^{(21)}\mathbf b
\end{array}
\right)
\end{eqnarray}
yields the correct solution ${\mathbf x}$ of system (\ref{lin}) in the rhs. Here $0_{M}$ is a zero column of $M$ entries. Substituting (\ref{lin}) for ${\mathbf b}$, we rewrite this equation as  
\begin{eqnarray}\label{UAx}
\left(
\begin{array}{cc}
U^{(11)}&U^{(12)}\cr
U^{(21)}& U^{(22)}
\end{array}
\right)
\left(
\begin{array}{c}
A \mathbf x\cr
0_M
\end{array}
\right)= \left(
\begin{array}{c}\mathbf x\cr
U^{(21)} A\mathbf x
\end{array}
\right).
\end{eqnarray}
It follows from (\ref{UAx}) that 
\begin{eqnarray}\label{Ux}
&&U^{(11)} A \mathbf x = \mathbf x.
\end{eqnarray}
Therefore 
\begin{eqnarray}\label{UA}
U^{(11)}=A^{-1}
\end{eqnarray}
or
\begin{eqnarray}\label{Ux4}
 U^{(11)}_{i j}  = A^{-1}_{ij} =\frac{(-1)^{i+j} M_{ji}}{\det\, A},
\end{eqnarray}
where we use the definition of the   inverse matrix $A^{-1}$ elements  in terms of the minors $M_{ij}$ of the matrix $A$,  and the minor $M_{ij}$ is the determinant of the matrix $A$  obtained by deleting the $i$th row and $j$th column of this matrix.
Since $U^{(11)}$ is a block of a unitary matrix, the vector norm of its column and rows can not exceed one. Then  relation (\ref{UA}) and formula (\ref{Ux4}) yield the following constraints on the rows   and columns 
of the matrix $A^{-1}$: 
\begin{eqnarray}\label{norm}
&&
\frac{1}{|\det A|}\sqrt{\sum_{i=1}^M  M_{ji}^2} =r_{0j}\le 1, \;\; j=1,\dots,M,\\\label{norm2}
&&
\frac{1}{|\det A|}\sqrt{\sum_{j=1}^M  M_{ji}^2} =r_{i0}\le 1, \;\; i=1,\dots,M.
\end{eqnarray}
The elements of all other blocks $U^{(12)}$, $U^{(21)}$  and $U^{(22)}$ must provide the  hermiticity of $U$: $U U^+=E_{M+N}$ 
(here and below $E_{K}$ is the 
$K\times K$ identity matrix).
Consequently, the block $U^{(12)}$ must be found from the equation 
\begin{eqnarray}\label{UU}
U^{(12)}(U^{(12)})^+ + U^{(11)} (U^{(11)})^+ = E_M.
\end{eqnarray}
The rows of the blocks $U^{(21)}$ and  $U^{(22)}$ can be found by the Gram-Schmidt orthogonalization algorithm. They satisfy the equations
\begin{eqnarray}
U^{(21)} (A^{-1})^+ + U^{(22)} (U^{(12)})^+  = 0,\;\;
U^{(21)} (U^{(21)})^+ + U^{(22)} (U^{(22)})^+  = E_M.\;\;
\end{eqnarray}

\subsection{Decreasing  dimensionality of  unitary transformation} 
\label{Section:lowdim}
The dimensionality of the unitary transformation can be reduced to  $M+1$ if we  calculate the needed elements of $\mathbf x$ one by one. 
This might be important for solving a system of linear equations via a minimal  quantum system.
To find the $x_k$ element of ${\mathbf x}=(x_1\dots x_M)$,  we 
introduce the unitary operator
\begin{eqnarray}\label{U0k}
U^{(k)}=\left(
\begin{array}{cc}
U^{(k;11)}&U^{(k;12)}\cr
U^{(k;21)}& U^{(k;22)}
\end{array}
\right)
\end{eqnarray}
and
consider the following equation:
\begin{eqnarray}\label{UAx2}
\left(
\begin{array}{cc}
U^{(k;11)}&U^{(k;12)}\cr
U^{(k;21)}& U^{(k;22)}
\end{array}
\right)
\left(
\begin{array}{c}A \mathbf x\cr
0
\end{array}
\right)= \left(
\begin{array}{c}
x_k\cr
U^{(k;21)} A\mathbf x
\end{array}
\right), 
\end{eqnarray}
which differs from eq. (\ref{UAx}) by the structure of the column in the rhs and by the dimensionalities of the blocks $U^{k;ij}$. Now  
 $U^{(k;11)}$ is a row  of $M$ elements, $U^{(k;12)}$ is a scalar,   
$U^{(k;21)}$ is an $M\times M$ matrix,  and  $U^{(k;22)}$ is a column of $M$ elements, so that  $U^{(k)}$ (\ref{U0k}) is an $(M+1)\times (M+1)$ matrix. 
It follows from Eq.(\ref{UAx2}):
\begin{eqnarray}\label{Ux3}
\sum_{i,j} U^{(k;11)}_{i} A_{ij} x_j = x_k,
\end{eqnarray}
or
\begin{eqnarray}\label{Ux1}
\sum_{i} U^{(k;11)}_{ i} A_{ij} =\delta_{kj},
\end{eqnarray}
where $\delta_{kj}$ is the Kronecker symbol.
Therefore
\begin{eqnarray}\label{Ux42}
 U^{(k;11)}_{ j}  = A^{-1}_{kj} =\frac{(-1)^{k+j} M_{j k}}{\det\, A}.
\end{eqnarray}
Thus, if we need to find only one component $x_k$, then  conditions (\ref{norm}) and (\ref{norm2}) reduce to a single inequality:
\begin{eqnarray}\label{r}
r_{k0}\le 1.
\end{eqnarray}
Instead of  (\ref{UU}), we have a scalar equation for the element  $U^{(k;12)}$:
\begin{eqnarray}\label{UU2}
(U^{(k;12)})^2 + \sum_{i=1}^M (U^{(k;11)}_i)^2 = 1 \;\;\Rightarrow\;\; U^{(k;12)} = \sqrt{1-\sum_{i=1}^M \left(\frac{M_{ik}}{\det \,A}\right)^2}.
\end{eqnarray}
Other rows of $U^{(k)}$ can be constructed by the Gram-Schmidt orthogonalization algorithm to satisfy the condition $U^{(k)}(U^{(k)})^T=E_{M+1}$.


Of course, if we need to find all the elements of ${\mathbf x}$, then we  have to construct $M$ unitary transformations $U^{(k)}$, $k=1,\dots, M$. Then, (\ref{r}) must hold for all $k=1,\dots,M$.

\section{Solving algebraic systems on superconducting quantum processor of IBM Quantum Experience}
\label{Section:IBM}
According to  Solovay-Kitaev theorem \cite{NCh},
any unitary operator can be approximated  by a superposition of CNOTs and single-qubit operations. 
Here we show how the unitary operators solving systems of linear algebraic equations can be exactly simulated using CNOT and single-qubit rotations. We emphasize that we are interested in such operators that commute with $I_z$, $[U,I_z]=0$. Together with one-excitation initial state, this requirement reduces the set of basis states involved into the process so that the quantum system evolves in the one-excitation state subspace.  

\subsection{Family of unitary transformations commuting with $I_z$}

We denote the CNOT between the $i$th and $j$th qubits with control qubit $i$ as $C_{ij}$. 
It can be  written in the basis of
\begin{eqnarray}
|0\rangle,\;\;|i\rangle,\;\;|j\rangle,\;\;|ij\rangle,
\end{eqnarray}
corresponding to the $i$th and $j$th excited spins:
\begin{eqnarray}
C_{ij}=\left(\begin{array}{cccc}
1&0&0&0\cr
0&1&0&0\cr
0&0&0&1\cr
0&0&1&0
\end{array}\right).
\end{eqnarray}
We introduce also the  one-qubit rotations
\begin{eqnarray}
R_{\alpha}(\phi)=\exp(i \phi I_{\alpha i}),\;\;\alpha =x,y,z,\;\;i=1,2,3,
\end{eqnarray}
where $\sigma_\alpha$, $\alpha =x,y,z$, are the Pauli matrices.
The 2-parametric unitary transformation of the $i$th spin reads
\begin{eqnarray}\label{Ri}
R_i(\alpha,\beta)=R_{zi}(\beta) R_{yi}(\alpha)R_{zi}(-\beta).
\end{eqnarray}
Now we can write a family of   unitary  transformations commuting with $I_z = \sum_i I_{zi}$:
\begin{eqnarray}\label{CC}
U_{ij}(\alpha,\beta)=C_{ij} R_{i}(\alpha,\beta) C_{ji} R_{i}^+(\alpha,\beta)C_{ij}.
\end{eqnarray}
In $U_{ij}$, the first index corresponds to the rotated qubit, and the second index corresponds to the qubit coupled with the rotated one by three CNOTs. 
This family can be extended by adding the $z$-rotation $R_{zi}$ of any qubit.  

\begin{figure*}
\epsfig{file=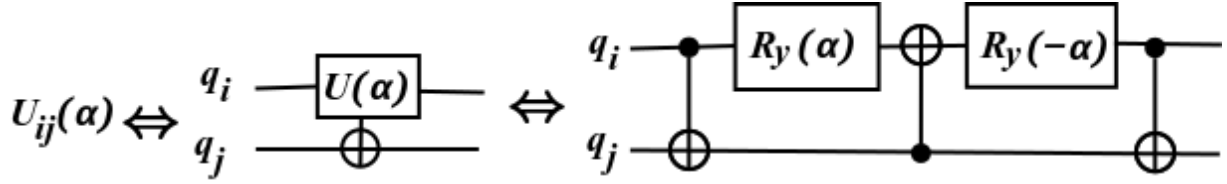,
scale=1,angle=0
} 
\caption{ The scheme of the unitary transformation  $U_{ij}(\alpha) \equiv U_{ij}(\alpha,0)$ which entangles the qubits $q_i$ and $q_j$. We don't  write the subscript $i$ in the $y$-rotation  operator $R_y$, and $R_y(\alpha)\equiv R_y(\alpha,0)$.}
  \label{Fig:Uij} 
\end{figure*}

For simplicity, hereafter  
 we consider  real matrices  $A$ and column ${\mathbf b}$. In this case, we  can put $\beta=0$ in the operators  $U_{ij}$,  $U_{ij}(\alpha,0) \equiv U_{ij}(\alpha)$.
The scheme of such operator $U_{ij}$ is  shown in Fig.\ref{Fig:Uij}, where we omit the subscript $i$ in the operator of $y$-rotation of the $i$th spin 
and put $R_{y}(\alpha) \equiv R_{y}(\alpha,0)$.

\subsection{Three-qubit quantum scheme for solving  system of two linear equations.}

We show that a linear system of two equations with real $A$ and ${\mathbf b}$,
\begin{eqnarray}\label{lin2}
A=\left(
\begin{array}{cc}
a_{11}&a_{12}\cr
a_{21}&a_{22}
\end{array}
\right),\;\;{\mathbf b} = \left(
\begin{array}{c}
b_1\cr
b_2
\end{array}
\right),
\end{eqnarray}
can be solved using a three-qubit quantum system.

\subsubsection{Initialization of the vector $b$}  
\begin{figure*}
\epsfig{file=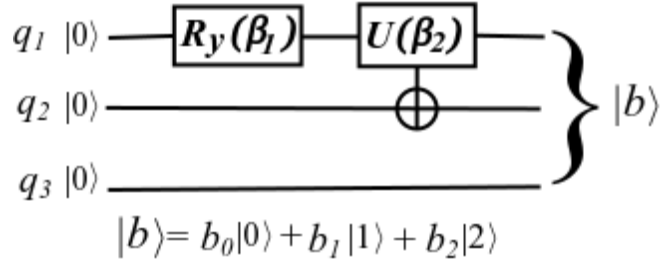,
scale=1,angle=0
} 
\caption{Initialization of the input vector ${\mathbf b}$. The third q-bit is not used.}
  \label{Fig:b} 
\end{figure*}

Representing  ${\mathbf b}$ as a quantum state is the first step of the protocol.
In our example, we consider such ${\mathbf b}$ that
 $|b_1|^2+|b_2|^2<1$.
In this case, we can encode the vector $b=(b_1\;\;b_2)^T$ into the following pure state with single excitation:
\begin{eqnarray}\label{b2}
&&
|\Psi\rangle_b =b_0|0\rangle+b_1|1\rangle + b_2 |2\rangle,\\\label{normb}
&&
b_0=\sqrt{1-|b_1|^2-|b_2|^2 }.
\end{eqnarray}
To produce this state we apply the unitary operator 
\begin{eqnarray}\label{Uop}
U_b(\beta_1,\beta_2)=U_{12}(\beta_2)R_y(\beta_1)
\end{eqnarray}
 to the  ground state $|0\rangle$  obtaining
\begin{eqnarray}
&&
|\Psi_b\rangle=U_{12}(\beta_2)R_1(\beta_1) |0\rangle = \\\nonumber
&&
\cos\frac{\beta_1}{2} |0\rangle
+  \sin\beta_2\sin \frac{\beta_1}{2} |1\rangle
-\cos \beta_2 \sin \frac{\beta_1}{2}|2\rangle.
\end{eqnarray}
The scheme of $U_{12}$  is given in Fig.\ref{Fig:b}.
Now we require
\begin{eqnarray}
\label{b1b2}
\left\{\begin{array}{l}
  \sin\beta_2\sin \frac{\beta_1}{2}=b_1\cr
-\cos \beta_2   \sin \frac{\beta_1}{2} =b_2
\end{array}\right.\;\;\Rightarrow \;\;
\left\{\begin{array}{l}\displaystyle
  \tan \beta_2 =-\frac{b_1}{b_2}\cr\displaystyle
  \sin\frac{\beta_1}{2} =\pm \sqrt{b_1^2+b_2^2}.
\end{array}\right.
\end{eqnarray}
Thus, for a given $b_1$ and $b_2$, we can find $\beta_i$, $i=1,2$.

\subsubsection{Unitary operators solving  algebraic system}
\label{Section:123}
\begin{figure*}
\epsfig{file=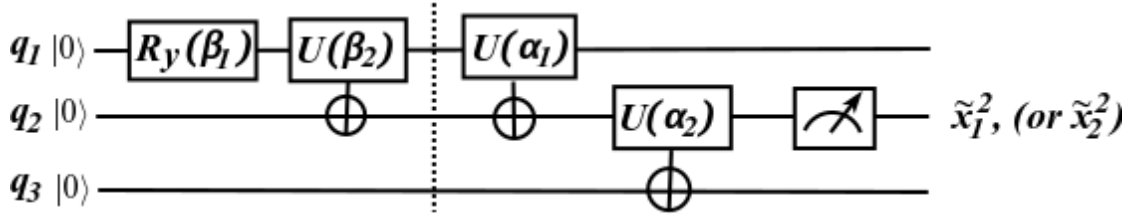,
scale=1,angle=0}
\caption{Complete scheme for solving a system of two equations. The vertical line separates the block initializing the input column ${\mathbf b}$. The structure of the blocks  $U(\alpha_i)$, $i=1,2$, is shown in Fig.\ref{Fig:Uij}. 
As the  result of  measurement, we obtain $\tilde x_1^2$ (if $\alpha_i$, $i=1,2$,  are defined in (\ref{DD1x})), or $\tilde x_2^2$ (if $\alpha_i$, $i=1,2$,  are defined in (\ref{DD2y})) as the probability for the measured qubit to be in the state $|1\rangle$. { In the  case of the  ideal quantum gates, $\tilde x_i^2 \equiv x_i^2$, $i=1,2$. }}
  \label{Fig:U2} 
\end{figure*}

Let us introduce the
two-parametric unitary transformation 
\begin{eqnarray}\label{U}
U_{123}(\alpha_1,\alpha_2)=U_{23}(\alpha_2) U_{12}(\alpha_1)
\end{eqnarray}
and apply this transformation to the state $|\Psi_b\rangle$.
The scheme of this operation together with the initialization of  the input data ${\mathbf b}$ is shown in Fig.\ref{Fig:U2}.
Since, according to linear equation (\ref{lin}),
\begin{eqnarray}\label{b}
b_i=a_{i1} x_1 + a_{i2} x_2,\;\;i=1,2,
\end{eqnarray}
we obtain
the three-qubit state $\Psi_U$ having the following structure
\begin{eqnarray}\label{DD}
|\Psi_U\rangle= U_{123} |\Psi_b\rangle=
(x_1 D_{11} +
x_2 D_{12}) |1\rangle+
(x_1 D_{21}+
x_2 D_{22})|2\rangle+
(x_1 D_{31} +
x_2 D_{32})|3\rangle,
\end{eqnarray}
where $D_{ij}$ are the following  expressions of $a_{ij}$ and $\alpha_i$:
\begin{eqnarray}
&&
D_{11}=a_{21}\cos \alpha_1-a_{11}\sin \alpha_1,\;\;
D_{12}=a_{22}\cos \alpha_1 -a_{12}\sin \alpha_1,\\\nonumber
&&
D_{21}=-D_1 \sin \alpha_2 ,\;\;
D_{22}=- D_2\sin \alpha_2 ,\\\nonumber
&& 
D_{31}=D_1\cos \alpha_2 ,\;\;
D_{32}=D_2\cos \alpha_2 ,\\\nonumber
&&
D_1 = a_{11}\cos \alpha_1 +a_{21}\sin \alpha_1,\;\;
D_2= a_{12}\cos \alpha_1 +a_{22}\sin \alpha_1.
\end{eqnarray}
With two arbitrary parameters $\alpha_i$, $i=1,2$, 
we can set a desired value to two of the coefficients $D_{ij}$ in (\ref{DD}).
{As a result, $x_i$, $i=1,2$, appear as the probability amplitudes in front of  $|2\rangle$ in the state $|\Psi_U\rangle$:}
\begin{eqnarray}\label{DD1}\label{DD1x}
{\mbox{if}}\;\;
\left\{
\begin{array}{l}
D_{21}=1\cr
D_{22}=0 
\end{array}
\right.
\;\;
\Rightarrow\;\;\;
\left\{
\begin{array}{l}\displaystyle
\tan\alpha_1 =-\frac{a_{12}}{a_{22}}\cr\displaystyle
\sin\alpha_2=\pm\frac{\sqrt{a_{22}^2 + a_{12}^2}}{\det\;A}
\end{array}
\right.
\;\;\Rightarrow \;\;
\langle 2|\Psi_U\rangle =x_1,\\\label{DD2}\label{DD2y}
{\mbox{if}} \;\;\;
\left\{
\begin{array}{l}
D_{21}=0\cr
D_{22}=1 
\end{array}
\right.
\;\;
\Rightarrow\;\;\;
\left\{
\begin{array}{l}\displaystyle
\tan\alpha_1 =-\frac{a_{11}}{a_{21}}\cr\displaystyle
\sin\alpha_2=\pm\frac{\sqrt{a_{11}^2 + a_{21}^2}}{\det\;A}
\end{array}
\right.
\;\;\Rightarrow \;\;
\langle 2|\Psi_U\rangle =x_2.
\end{eqnarray}
In this way, we can obtain the  value of either  $x_1$ or $x_2$ using different values of the parameters $\alpha_i$, $i=1,2$, in the unitary transformation. In both cases, the variable $x_i$ appears as a probability amplitude for the state transfer  $|\Psi_U\rangle \to |2\rangle$. { Due to the probabilistic method of obtaining the result, we measure $\tilde x_i^2$ (remember that $x_i$, $i=1,2$, are real)  which doesn't equal  $ x_i^2$ due to the imperfections of quantum gates, similar to Sec.\ref{Section:123}. }

\subsubsection{Example}
\label{Section:example}
We consider  the following $A$ and ${\mathbf b}$:
\begin{eqnarray}\label{lin3}
A=\left(
\begin{array}{cc}
-1.8&0.6\cr
-0.4 & 1.4
\end{array}
\right)
,\;\;{\mathbf b}= \left(
\begin{array}{c}
-0.6 \cr 0.8
\end{array}
\right).
\end{eqnarray}
For such matrix $A$, 
 condition (\ref{r}) holds for both rows of $A^{-1}$.

First, we define the parameters $\beta_i$, $i=1,2$, in the unitary transformation $U_b$ (\ref{Uop}). 
Since $b_1^2+b_2^2=1$ in this case, we set $\beta_1=-\pi$ which yields  $b_0=0$. Then  the second of equations (\ref{b1b2}) holds, while the first one yields
\begin{eqnarray}
\beta_2=-\arctan \frac{b_1}{b_2}=0.64350.
\end{eqnarray}

Next, we find the parameters $\alpha_i$, $i=1,2$,  in $U_{123}$ (\ref{U}).  
Formulas  (\ref{DD1})  
yield: 
\begin{eqnarray}\label{par1}
\alpha_1=2.73670, \;\;\;\alpha_2= 5.55160 \;\;\;
{\Rightarrow} \;\;\;
x_1=\langle 2|\Psi_U\rangle =0.5789.
\end{eqnarray}
Formulas (\ref{DD2}) yield:
\begin{eqnarray}\label{par2}
\alpha_1=1.78947,\;\;\; \alpha_2=5.34119 \;\;\;
{\Rightarrow} \;\;\;
x_2=\langle 2|\Psi_U\rangle =0.7368.
\end{eqnarray}
Of course, in both (\ref{par1}) and (\ref{par2}), $\alpha_i$, $i=1,2$, are not unique.

\subsection{Four qubit quantum scheme and system of three linear equations}
\label{Section:3eq}
\subsubsection{Initialization of ${\mathbf b}$ and construction of unitary transformations }
\label{Section:3eq2}
\begin{figure*}
\epsfig{file=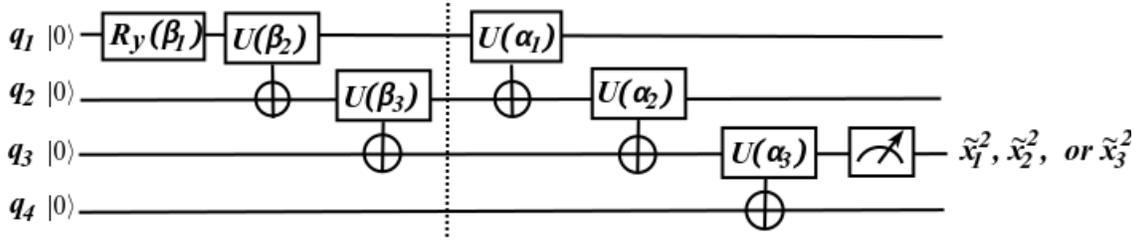,
scale=0.8,angle=0}
\caption{Complete scheme for solving a system of three equations. The vertical line separates the block initializing the input column ${\mathbf b}$. The structure of the blocks  $U(\beta_i)$ and $U(\alpha_i)$ is shown in Fig.\ref{Fig:Uij}. For the $\alpha_i$ satisfying one of systems   (\ref{s1}), (\ref{s2}) or (\ref{s3}),  we obtain one of the quantities {
$\tilde x_1^2$, $\tilde x_2^2$ or $\tilde x_3^2$  as the  probability for the measured qubit to be in the state $|1\rangle$.  In the  case of  ideal quantum gates, $\tilde x_i^2 \equiv x_i^2$, $i=1,2,3$. }}
  \label{Fig:U3} 
\end{figure*}

We need the set of three unitary transformations $R_{i}$ (\ref{Ri}), $i=1,2,3$,
and three unitary  transformations commuting with $I_z$: $U_{12}$, $U_{23}$ and $U_{34}$.
To initialize the input vector $|\Psi_b\rangle$, we apply the transformation 
\begin{eqnarray}\label{Ub3}
U_{b}(\beta_1,\beta_2,\beta_3) = U_{23}(\beta_3)U_{12}(\beta_2) R_1(\beta_1)
\end{eqnarray}
to the ground state: 
\begin{eqnarray}
|\Psi_b\rangle = U_b |0\rangle = \sum_{i=0}^3 D^{(b)}_i|i\rangle,
\end{eqnarray}
where $D^{(b)}_i$, $i=0,\dots,3$ are the known expressions of $\beta_i$. 
Then the system of equations for $\beta_i$, $i=1,2,3$, reads
\begin{eqnarray}\label{N4beta}
D^{(b)}_i=b_i,\;\;i=1,2,3.
\end{eqnarray}
Next, to find $x_i$, $i=1,2,3$, we apply the transformation
\begin{eqnarray}\label{U3}
U_{1234}(\alpha_1,\alpha_2,\alpha_3)=U_{34}(\alpha_3)U_{23}(\alpha_2) U_{12}(\alpha_1)
\end{eqnarray}
to 
$|\Psi_b\rangle$ obtaining $|\Psi_U\rangle$:
\begin{eqnarray}
|\Psi_U\rangle = U_{1234} |\Psi_b\rangle.
\end{eqnarray}
The state $|\Psi_U\rangle$ is a superposition of states $|n\rangle$, $n=0,\dots,4$, where the probability amplitude of, for instance, the state transfer 
$|\Psi_U\rangle \to |3\rangle$  reads
\begin{eqnarray}\label{p}
\langle 3|\Psi\rangle = D_1 x_1 + D_2 x_2 + D_3 x_3,
\end{eqnarray}
where $D_i$, $i=1,2,3$, are the known functions of $\alpha_i$, $i=1,2,3$. We do not represent the explicit expressions for $D_i$.  Three parameters $\alpha_i$, $i=1,2,3$, can control three functions $D_i$, $i=1,2,3$. {Thus, $x_i$, $i=1,2,3$, appear as the probability amplitudes in front of  $|3\rangle$ in the state $|\Psi_U\rangle$:}  
\begin{eqnarray}\label{s1}
&&{\mbox{if}}\;\;\;D_1=1,\;\;D_2=0,\;\;D_3=0 \;\; \Rightarrow\;\;\langle 3|\Psi_U\rangle = x_1, \\\label{s2}
&&{\mbox{if}}\;\;\;D_1=0,\;\;D_2=1,\;\;D_3=0  \;\; \Rightarrow\;\;\langle 3|\Psi_U\rangle = x_2, \\\label{s3}
&&{\mbox{if}}\;\;\;D_1=0,\;\;D_2=0,\;\;D_3=1  \;\; \Rightarrow\;\;\langle 3|\Psi_U\rangle = x_3.
\end{eqnarray}
The scheme of this protocol is shown in Fig.\ref{Fig:U3}.
{ Systems (\ref{N4beta}) for $\beta_i$, $i=1,2,3$ and (\ref{s1})-(\ref{s3}) for $\alpha_i$, $i=1,2,3$, are rather cumbersome, and we do not provide general solution for them. Instead, we give a particular solution for the fixed $A$ and $\mathbf b$ in the example below.}

\subsubsection{Example}
\label{Section:ex2}
We consider the following  $A$ and ${\mathbf b}$: 
\begin{eqnarray}\label{AN3}
A=\left(\begin{array}{ccc}
0.9&-0.6&-1.8\cr
1.6&-0.5&-0.6\cr
0.8&-1.4&-0.5
\end{array}\right),\;\;\;{\mathbf b}=\left(\begin{array}{c}
-0.5\cr
0.7\cr
-0.5
\end{array}\right).
\end{eqnarray}
In this case, condition (\ref{r}) holds for all the columns of $A^{-1}$ and $\sqrt{\sum_{i=1}^3b_i^2}=0.77<1$.
System (\ref{N4beta}) yields the following particular  values for the parameters $\beta_i$, $i=1,2,3$, in the transformation  $U_b$ (\ref{Ub3}):
\begin{eqnarray}
\beta_1=2.94126,\;\;\;
\beta_2=3.66810,\;\;\;
\beta_3=4.09214.
\end{eqnarray}
In turn, systems (\ref{s1})-(\ref{s2}) yield the following particular values for the parameters $\alpha_i$ ,$ i=1,2,3$ in $U_{1234}$ (\ref{U3}):
\begin{eqnarray}\label{xxx}
&&\alpha_1=1.83056, \;\;\; \alpha_2=6.05229, \;\;\; \alpha_3=5.13645 \;\;\Rightarrow \;\;\;\langle 3|\Psi_U\rangle = x_1=0.8185,\\\nonumber
&&\alpha_1=1.25816 , \;\;\; \alpha_2=5.13077, \;\;\;\alpha_3=4.85991 \;\;\Rightarrow \;\;\;\langle 3|\Psi_U\rangle = x_2=0.6578,\\\nonumber
&&\alpha_1=5.88224 , \;\;\; \alpha_2=2.89173, \;\;\;\alpha_3=5.36144 \;\;\Rightarrow \;\;\;\langle 3|\Psi_U\rangle = x_3=0.4677.
\end{eqnarray}

\subsubsection{Simulation on 5-qubit superconducting quantum processor of IBM Quantum Experience }

Now we discuss the realization of the protocol for solving a system of three equations on the quantum processor of IBM Quantum Experience, see Fig.\ref{Fig:U3}.
The solution $x_i$ of the equation can be registered as the result of measurement on a particular qubit of this processor {($q_3$ in Fig.\ref{Fig:U3}). Measuring yields the state $|1\rangle$ with the probability  $x_i^2$ in the ideal case. Therefore, the probabilistic result is $x_i^2$  rather then $x_i$. In reality, due to the imperfections of quantum gates and final number  (equal to 1024) of protocol running, we measure  $\tilde x_i^2$, $i=1,2,3$, which defer from the ideal values  $x_i^2$}. 

We compare the quantities $\tilde x_i^2$, $i=1,2,3$,  calculated using the above quantum processor with the true values  of the variables $x_i^2$, $i=1,2,3$,  obtained via the classical methods. 
In all calculations, we average the result over four series of measurements,  each series includes 1024 independent  runs of the algorithm. 

\begin{figure*}
\epsfig{file=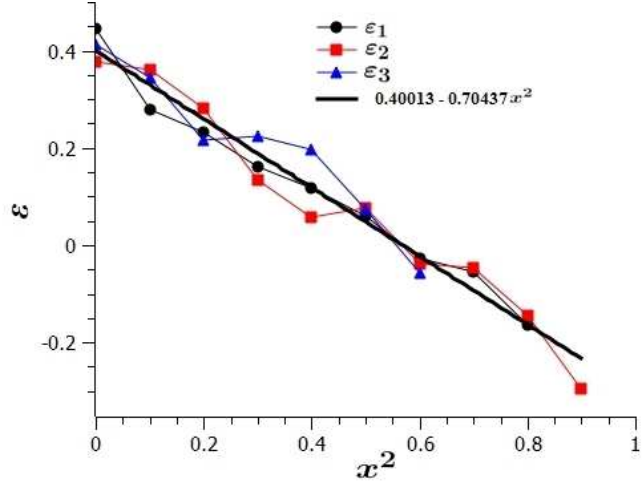
,scale=0.5,angle=0
} 
\caption{The errors  $\varepsilon_i \equiv \tilde x_i^2 - x_i^2$, $i=1,2,3$, as functions of $x_i^2$ for  sets of $x_i^2$ (\ref{x1})-(\ref{x3}). In Figs. \ref{Fig:x}- \ref{Fig:random}, circles, squares and triangles correspond, respectively, to
$x_1^2$, $x_2^2$ and $x_3^2$. All points are settled around the line (\ref{line}). In Figs \ref{Fig:x}-\ref{Fig:xcr},  the linear system with the matrix $A$ given in (\ref{AN3}) is considered}
  \label{Fig:x} 
\end{figure*}

We use  matrix $A$ (\ref{AN3}) considered in Sec.\ref{Section:ex2}, while the input vector ${\mathbf b}$ varies.
Only one entree $x_i$ of ${\mathbf x}$ can be measured in our protocol and $x_i^2$ can not exceed 1 since it is the  probability of a  certain state. Therefore $0\le x_i^2\le 1$. To characterize the accuracy of  calculations, 
we take a set of values multiple of $0.1$ for each variable $x_i$:
\begin{eqnarray}
\label{x1}
x_1^2=0.1 n, \;\;n=0,\dots,8, \\\label{x2}
x_2^2=0.1 n, \;\;n=0,\dots,9, \\\label{x3}
x_3^2=0.1 n, \;\;n=0,\dots,6.
\end{eqnarray}
{ The upper boundary for each $x_i^2$ in (\ref{x1})-(\ref{x3}) depends on a particular choice of the matrix $A$ (Eq.(\ref{AN3}) in our case).  
For each  value of $x_1^2$, $x_2^2$ or $x_3^2$, we fix the values of two other variables in a random way and find the appropriate vector ${\mathbf b}$ using Eq.(\ref{lin}).
 Thus, we construct three sets of vectors ${\mathbf b}_i$, $i=1,2,3$, corresponding to  sets (\ref{x1})-(\ref{x3}). Next, for  the found sets 
 ${\mathbf b}_i$ and matrix $A$,
we perform  the protocol, presented in Sec.\ref{Section:3eq}, on a quantum processor using the above-described averaging procedure and find the appropriate values $\tilde x_i$, thus constructing three sets of quantities $\tilde x_i$, $i=1,2,3$. Schematically, these steps can be represented as the following maps:
\begin{eqnarray}
x_i^2 \; \to \; {\mathbf b}_i \; \to \; \tilde x_i,\;\;i=1,2,3.
\end{eqnarray}
In the ideal case,  $x_i^2 \equiv \tilde x_i^2$. But this equality doesn't hold in reality, and we introduce
the error $\varepsilon_i$, 
\begin{eqnarray}\label{varepsilon}
\varepsilon_i = \tilde x^2_i - x^2_i,
\end{eqnarray}
to characterize the deviation of the measured values from the true ones. The error $\varepsilon_i$ is  shown in Fig.\ref{Fig:x}}, where  circles, squares and triangles correspond, respectively, to
$x_1^2$, $x_2^2$ and $x_3^2$. This figure shows that, instead of an identical zero values $\varepsilon_i$, $i=1,2,3$, expected in the ideal case, we have a set of points significantly  different from zero line. However, all these points are settled around the straight line constructed  by the least-square method and shown in the same Fig.\ref{Fig:x}:
\begin{eqnarray}\label{line}
\varepsilon
=0.40013-0.70437 x^2.
\end{eqnarray}
The character of the point distribution in Fig.\ref{Fig:x} prompts us to consider the line shown in this figure as the correction function which must be subtracted from the result calculated on the quantum processor. {
In other words, for any  measured value $\tilde x_i^2$  we introduce the quantity $X_i$  by the formular
\begin{eqnarray}\label{corrF}
X_i(x_i)=\tilde x_i^2 - \varepsilon
\end{eqnarray}
and consider $X_i$ as the  result of execution  of the protocol on the quantum processor. }

\begin{figure*}
\epsfig{file=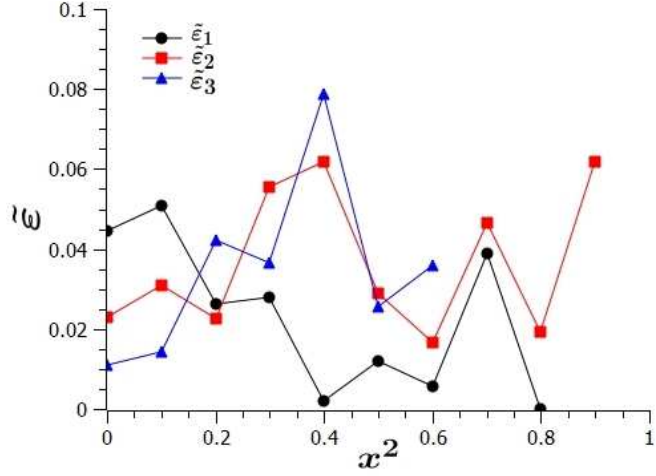,
scale=0.5,angle=0
} 
\caption{The corrected errors  $\tilde \varepsilon_i$, $i=1,2,3$, as functions of $x_i^2$
for  sets of $x_i^2$ (\ref{x1})-(\ref{x3}); 
 $\tilde\varepsilon_i < 0.08$.}
  \label{Fig:xc} 
\end{figure*}

\begin{figure*}
\epsfig{file=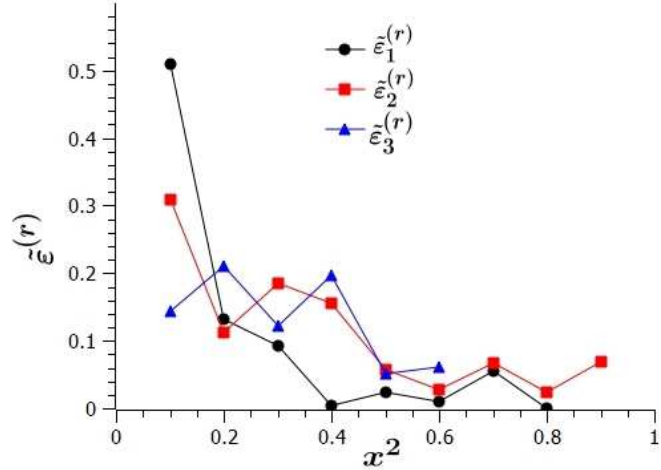,
scale=0.5,angle=0
} 
\caption{The relative  errors  $\tilde \varepsilon^{(r)}_i$, $i=1,2,3$, as functions of $x_i^2$
for  sets of $x_i^2$ (\ref{x1})-(\ref{x3}). 
$\tilde \varepsilon^{(r)}_i$ is large for $x_i^2\lesssim 0.2$. The errors $\tilde \varepsilon^{(r)}_i(0)$, $i=1,2,3$,  tend to infinity and are not shown here. }
  \label{Fig:xcr} 
\end{figure*}

The errors 
\begin{eqnarray}\label{err}
\tilde \varepsilon_i = X_i(x_i)- x_i^2, \;\;\; i=1,2,3
\end{eqnarray}
as functions of $x_i^2$  are depicted in Fig.\ref{Fig:xc}. We see that the absolute values of these errors do not exceed 0.08.
However, the relative error  
\begin{eqnarray}\label{rel}
\tilde \varepsilon^{(r)}_i = \frac{\tilde \varepsilon_i}{x_i^2}
\end{eqnarray}
is significant for $x_i^2\lesssim 0.2$ as shown in Fig.\ref{Fig:xcr} (the errors $\tilde \varepsilon^{(r)}_i(0)$ tend to infinity and are not shown in this figure). Thus, the proposed algorithm for solving the systems of linear algebraic equations
gives reasonable results for $x_i^2\gtrsim 0.2$. 

{{}We underline a principal differences among the three introduced errors: $\varepsilon_i$ (Eq.(\ref{varepsilon}) and Fig.\ref{Fig:x}), $\tilde \varepsilon_i$ (Eq.(\ref{err}) and Fig.\ref{Fig:xc}) and $\tilde \varepsilon^{(r)}_i$ (Eq.(\ref{rel}) and Fig.\ref{Fig:xcr}). The error $\varepsilon_i$ indicates imperfections in  realization of quantum operations and measurements on  superconducting qubits. The second error $\tilde \varepsilon_i$ is significantly less than  $\varepsilon_i$, it shows that the measured   results $\tilde x_i$  can be corrected to obtain $X_i$ through formula
(\ref{corrF}) using a specially constructed correction function (\ref{line}). At last, the relative error  $\tilde \varepsilon^{(r)}_i$ shows that the  corrected result $X_i$ still is not reliable for small $x_i^2\lesssim 0.2$.}

\begin{figure*}
{\includegraphics[scale=0.5,angle=0]{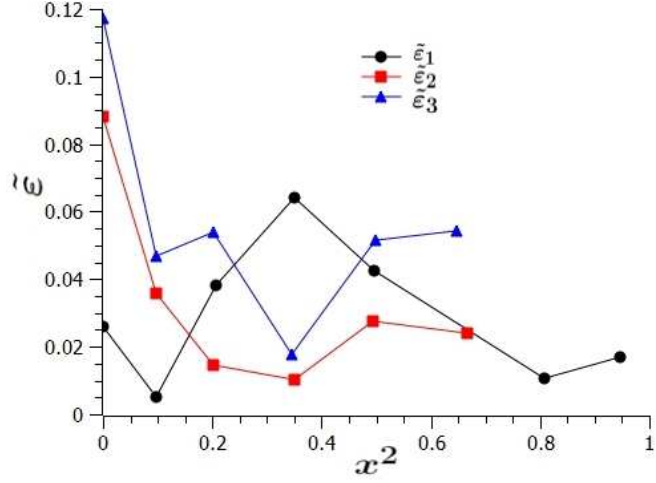
}} \\

\centerline{(a)}

{\includegraphics[scale=0.5,angle=0]{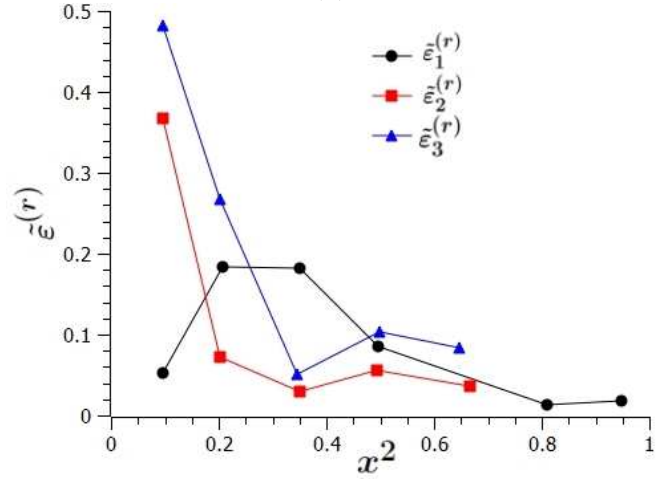
}}

\centerline{(b)}

\caption{The linear system with the matrix $A$ given in (\ref{A2}). The absolute $\tilde \varepsilon_i$ (a) and relative $\tilde \varepsilon^{(r)}_i$ (b) errors, $i=1,2,3$, as function of $x_i^2$.
These graphs are very similar to ones depicted in Figs.\ref{Fig:xc} and \ref{Fig:xcr}. The errors $\tilde \varepsilon^{(r)}_i(0)$, $i=1,2,3$,  tend to infinity and are not shown here. }
  \label{Fig:random} 
\end{figure*}

To demonstrate the usage of the introduced correction function (\ref{corrF}), we implement this function to correct the results obtained for a completely different matrix $A$ constructed  using the   pseudorandom number generator :
\begin{eqnarray}\label{A2}
A=\frac{2}{3}\left(\begin{array}{ccc}
-1.43& -1.10& -1.06\cr
0.818& 0.367& -1.42\cr 
-0.392& 1.60& -0.654
\end{array}
\right).
\end{eqnarray}
The absolute $\tilde \varepsilon_i$ and relative  $\tilde \varepsilon_i^{(r)}$ errors obtained using the protocol of this section with formulas (\ref{err}) and (\ref{rel}) are shown in Fig.\ref{Fig:random}. 
We notice that Fig.\ref{Fig:random}a and Fig.\ref{Fig:random}b  are very similar, respectively, to Fig.\ref{Fig:xc} and Fig.\ref{Fig:xcr}.

\section{Solving linear systems by means of spin-evolution operator}
\label{Section:field}
The evolution of the inhomogeneous spin-1/2 chain in the inhomogeneous external magnetic field
can be  a tool for solving the algebraic systems (\ref{lin}). In this case, the inverse of $A$ is implicitly encoded into the evolution operator, while the  input vector ${\mathbf b}$
must be encoded into the pure state of the spin chain. We consider the evolution  governed by the nearest-neighbor XX-Hamiltonian in the inhomogeneous magnetic field:
\begin{eqnarray}\label{H}
&&H=\sum_{i=1}^{M-1} d_i(I_{x,i}I_{x,(i+1)} + I_{y,i}I_{y,(i+1)}) + \sum_{i=1}^N (\omega-\omega_i) I_{z,i},\\\label{com}
&&[H,I_z]=0,\;\;I_z=\sum_i I_{z,i},
\end{eqnarray}
where $d_i$ are the coupling constants, $\omega_i$ are the Larmor frequencies $\omega_i=\gamma h_i$, $\gamma$ is the gyromagnetic ratio,  $h_i$, $i=1,\dots,M$, represent the inhomogeneous part of the external magnetic field. We set $\omega=\frac{1}{2}\sum_{i=1}^{N}\omega_i$. 
 The evolution of a spin system  reads
\begin{eqnarray}
V(t)=e^{-i H t},
\end{eqnarray}
so that $|\Psi(t)\rangle = V(t) |\Psi_b\rangle$. 
Using the proper values for the parameters $d_i$ and $\omega_i$ we can provide such evolution that, at some time instant, one of the amplitudes  of the pure state 
$|\Psi(t)\rangle$  equals one of the variables $x_i$, $i=1,2,3$, similar to the strategy of Sec.\ref{Section:3eq}.

Now we consider the particular example of a 4-qubit chain and adjust it for solving  a system of three equations considered in Sec.\ref{Section:ex2}. 
For this aim, we  find  the projection 
\begin{eqnarray}\label{proj}
\langle 3|\Psi\rangle=\sum_{i=1}^4 P_i x_i,
\end{eqnarray}
where $P_i$ depends on the time $t$ and the parameters of the Hamiltonian.  { To obtain the  value of a particular variable $x_k$ in (\ref{proj}), we solve the system 
\begin{eqnarray}
\label{P}
P_i=\delta_{ik},\;\;i=1,2,3,
\end{eqnarray}
for the parameters $d_i$ and $\omega_i$ (at a fixed time instant $t$). Then
\begin{eqnarray}\label{xi}
\langle 3|\Psi\rangle=x_k.
\end{eqnarray}}
We note that, unlike Sec.\ref{Section:ex2}, the coefficients $P_i$ are complex. Therefore we need six parameters to satisfy conditions (\ref{P}). Below we fix $d_1=1$ (which corresponds to the dimensionless time) and find the parameters $d_2$, $d_3$  and $\omega_i$, $i=1,\dots,4$, which satisfy (\ref{P}) at the  minimal possible time instant $t$. 
Doing this we  impose the constraints on the values of the parameters $d_i$ and $\omega_i$ caused by the nearest neighbor approximation (Hamiltonian (\ref{H}))
\begin{eqnarray}
0.1<d_i<2, \;\;i=1,2,\;\;\;-3<\omega_i<3,  \;\;i=1,\dots,4.
\end{eqnarray}
Results of direct computations are collected in Table \ref{Table:1}. 
\begin{table}
\begin{tabular}{|c|c|c|c|c|c|c|c|}
\hline
$x_i$&$d_2$&$d_3$&$\omega_1$&$\omega_2$&$\omega_3$&$\omega_4$&$t_{min}$\cr
\hline
$x_1$&1.92609& 1.10051&1.88349&-0.82883&-1.05897&0.37563&1.51485\cr
$x_2$&0.63225&1.59251&0.05200&2.89465&1.41259&-1.63479&2.05543\cr
$x_3$&1.52851&1.22234&1.74816&1.62240&2.16566&2.87055&3.64261\cr
\hline
\end{tabular}
\caption{\label{Table:1}Parameters of Hamiltonian (\ref{H}) solving system (\ref{P}) for different $i=1,2,3$.
}
\end{table}
For the initial state used in Sec.\ref{Section:ex2},
\begin{eqnarray}
|\Psi_0\rangle=b_0 |0\rangle+ b_1 |1\rangle+ b_2 |2\rangle+ b_3 |3\rangle,\;\;\sum_{i=0}^3 |b_i|^2=1 ,
\end{eqnarray}
with $b_1=b_3=-0.5$, $b_2=0.7$, 
we result in  $x_i$, $i=1,2,3$,  presented in (\ref{xxx}). We note that the time $t_{min}$ needed to perform the considered operation increases with the length of the chain, which is defined by the number of equations in the algebraic system. If we need to find all variables $x_i$, $i=1,\dots,K$, in a $K$-dimensional algebraic system, then the  required whole time interval equals the sum of the  time intervals needed for constructing each variable $x_i$. Therefore, if we need the shortest time interval, it might be more profitable to use $2 K$-qubit system and find the whole set of $x_i$, $i=1,\dots,K$, at ones, see the protocol in Sec.\ref{Section:encoding}.

\section{Conclusion}
\label{Section:conclusions}
We propose a protocol for solving  a system of linear algebraic equations using the  quantum system with the minimal number of qubits. The number of required qubits exceeds the number of equations in the linear system only by one. In this protocol, we use the properly constructed unitary transformation to find  each particular  variable $x_k$ leaving the other variables  undetermined.  Therefore, to completely solve a system of $M$ linear  equations we need to construct a set of $M$ different $(M+1)$-qubit unitary transformations. Constructing each unitary transformation requires calculating  $M$ minors of the system matrix $A$, which is equivalent to calculating $M$ elements of the inverse matrix $A^{-1}$. If we are interested in a particular $x_k$, then we need only one unitary transformation and 
other elements of $A^{-1}$ remains uncalculated.
Otherwise, if all $x_i$ must be calculated, then we have to find  all the elements of $A^{-1}$ using classical methods. Thus, in our protocol we combine the classical and quantum methods to reach the final purpose. However, been constructed, the unitary transformation(s) can be used further for  calculating 
$x_i$, $i=1,\dots,M$, for different ${\mathbf b}$. Formally, the above set of $M$ unitary  transformations in  the quantum algorithm replaces  the multiplication of $A^{-1}$ by a vector ${\mathbf b}$ in the classical case. 

We also study the implementation of the proposed protocol on the superconducting quantum processor of IBM Quantum Experience. In this case, we represent the needed unitary transformation as a superposition of the CNOTs (two-qubit operations) and one-spin rotations ($y$-axis rotations in the case of real $A$ and ${\mathbf b}$).  Analyzing solutions for systems of  three equations  obtained in this way, we found the accuracy available in such simulations. To increase the accuracy, the correction function is introduced, which must be subtracted from the result obtained via the quantum processor. Taking this function into account, we reduce the absolute error to $\sim 0.08$ and the relative error to $\sim 25 \%$ for large enough $x_i^2$
($x_i^2 \gtrsim 0.2$). The calculations for small $x_i\lesssim 0.2$, $i=1,\dots,M$, are still not reliable.

The advantage of such protocol is most obvious if we turn to the realization of the unitary transformation 
as an natural evolution operator for an $(M+1)$-qubit spin-1/2 chain governed, for instance,  by the nearest-neighbor XX Hamiltonian in the inhomogeneous magnetic field, see Sec.\ref{Section:field}. In this case, the required value of the variable $x_k$ appears as a probability amplitude of an excitation at a particular qubit. In this way, we solve a system of three equations using a spin chain of four qubits.

Authors acknowledge the use of the IBM Quantum Experience for this work.
This work is performed in accordance with the state task, state registration No. 0089-2019-0002. 
 The authors acknowledge the support from the Presidium of RAS, Program No.5 ''Photonic technologies in probing inhomogeneous media and biological objects''. 

\end{document}
 
 \bibitem{BBCVMSSSW}
 A.Barenco, Ch.H.Bennett,  R. Cleve, D.P.DiVincenzo, N.Margolus, P.Shor, T.Sleator, J.A.Smolin,
H.Weinfurter, 
Phys.Rev.A {\bf 52}, 3457 (1995)
 
 \bibitem{CWSCGZLLLP}
X.-D. Cai,  C. Weedbrook,  Z.-E. Su,  M.-C. Chen,  M. Gu, M.-J. Zhu,  L. Li,   N.-L. Liu, 
Ch.-Y. Lu, and J.-W. Pan, Phys.Rev.Lett. {\bf 110}, 230501 (2013) 
 
 \bibitem{BKRLDAW}
S.Barz, I.Kassal, M.Ringbauer, Y. O.Lipp, B.Daki\c,
A.Aspuru-Guzik, and P.Walther, Sci.Rep. {\bf 4}, 6115 (2014)
 
 \bibitem{WBL}
N. Wiebe,  D. Braun, 
and S. Lloyd, Phys.Rev.Lett. {\bf 109} (2012) 050505
  
\bibitem{ZSCXLG}
Y.Zheng,  Ch.Song, M.-Ch.Chen, B.Xia, W.Liu,  Q.Guo,  L.Zhang, 
D.Xu,  H.Deng,  K.Huang, Y.Wu,  Zh.Yan,  D.Zheng, L.Lu,  J.-W.Pan, 
H.Wang, Ch.-Y.Lu,  and X.Zhu, Phys.Rev.Lett. {\bf 118}, 210504 (2017)
 
 \bibitem{BWPRWL}
J.Biamonte, P. Wittek, N.Pancotti, P.Rebentrost, N.Wiebe, and S.Lloyd, Nature {\bf 549}, 195 (2017)

 \bibitem{HPXLG}
S.F. Huelga,  M. B. Plenio,  G.-Y. Xiang,  J. Li, and G.-C. Guo,
J. Opt. B: Quantum Semiclass. Opt. {\bf 7} (2005) S384

 \bibitem{QZACB}
 X.Qiang, X.Zhou, K.Aungskunsiri, H.Cable, and J. L O'Brien,
 Quantum Sci. Technol. {\bf 2} (2017) 045002

\bibitem{WY}
D. Wang, L. Ye, Int. J. Theor. Phys. {\bf 53} (2014) 350

\bibitem{FZ_2017}
E.B.Fel'dman, A.I.Zenchuk, JETP {\bf 125}, 1042 (2017)

\bibitem{Z_2018}
A.I.Zenchuk. Phys.Lett.A {\bf 382} (2018) 3244

 \bibitem{LP}
 A. Luis, J. Pe\u rina, Phys.Rev.A {\bf 54}, 4564 (1996)
 
 \bibitem{BACS}
 D.W.Berry, G. Ahokas, R.Cleve, B.C.Sanders, Commun. Math. Phys. {\bf 270}, 359 (2007)

 \bibitem{Kraus}
 K. Kraus, States, Effects and Operations (Spring-Verlag,
Berlin, 1983)

 \bibitem{FL}
 E. B. Fel'dman, S. Lacelle, Chem. Phys. Lett. {\bf 253}, 27
(1996)

\bibitem{BZ_2016}
 G. Bochkin and  A. Zenchuk, Quant. Inf. Comp., {\bf 16} 1349 (2016) 

\end{thebibliography}

\end{document}

 \bibitem{BFZ_2018} G.A.Bochkin, E.B.Fel'dman and A.I.Zenchuk,  Quantum Inf. Process. {\bf 17}, 218 (2018) 

\bibitem{Z_2018}
A.I.Zenchuk. Phys.Lett.A, accepted

\bibitem{Bose}
S. Bose, Phys. Rev. Lett. {\bf   91}, 207901 (2003)

\bibitem{CDEL}
 M.Christandl, N.Datta, A.Ekert, and A.J.Landahl, Phys.Rev.Lett. {\bf   92}, 187902 (2004)

\bibitem{ACDE}
 C.Albanese, M.Christandl, N.Datta, and A.Ekert, Phys.Rev.Lett. {\bf   93}, 230502 (2004)

\bibitem{KS}
 P.Karbach and J.Stolze, Phys.Rev.A {\bf   72}, 030301(R) (2005)

\bibitem{GKMT}
 G.Gualdi, V.Kostak, I.Marzoli, and P.Tombesi, Phys.Rev. A {\bf   78}, 022325 (2008)
 
\bibitem{WLKGGB}
A.W\'ojcik, T.Luczak, P.Kurzy\'nski, A.Grudka, T.Gdala, and M.Bednarska
Phys. Rev. A {\bf   72}, 034303 (2005)

 \bibitem{BACVV2010}
 L. Banchi, T. J. G. Apollaro,  A. Cuccoli,  R. Vaia, and P. Verrucchi, 
 Phys.Rev.A {\bf 82}, 052321 (2010)

\bibitem{ZO}
A. Zwick and O. Osenda, J. Phys. A: Math. Theor. {\bf 44}, (2011) 105302.

 \bibitem{BACVV2011}
 L. Banchi, T. J. G. Apollaro, A. Cuccoli, R. Vaia
and P. Verrucchi,
 New J. Phys. {\bf 13}, 123006 (2011) 

\bibitem{ABCVV}
T. J. G. Apollaro, L. Banchi, A. Cuccoli, R. Vaia, and
P. Verrucchi, Phys. Rev. A {\bf 85} (2012), 052319

 \bibitem{SAOZ}
 J.Stolze, G. A. \'Alvarez,
O. Osenda, and  A. Zwick in
{\it Quantum State Transfer and Network Engineering.
Quantum Science and Technology},
ed. by  G.M.Nikolopoulos and I.Jex, Springer Berlin Heidelberg, Berlin, p.149  (2014)

\bibitem{PBGWK2}
N.A.Peters, J.T.Barreiro,  M.E.Goggin, T.-C.Wei,  and P.G.Kwiat, Phys.Rev.Lett. {\bf   94}, 
150502 (2005) 

\bibitem{PBGWK}
N.A.Peters, J.T.Barreiro, M.E.Goggin, T.-C.Wei, and P.G.Kwiat in {\it Quantum
Communications and Quantum Imaging III}, ed. R.E.Meyers,
Ya.Shih, Proc. of SPIE {\bf   5893} (SPIE, Bellingham, WA, 2005) 

\bibitem{XLYG}
G.Y. Xiang, J.Li, B.Yu, and G.C.Guo
Phys. Rev. A {\bf   72}, 012315  (2005)

\bibitem{Z_2014}
A.I.Zenchuk, 
Phys. Rev. A {\bf 90}, 052302(13) (2014)

\bibitem{BZ_2015}
G. A. Bochkin and A. I. Zenchuk, 
Phys.Rev.A 91, 062326(11) (2015)

\bibitem{Werner}
R. F. Werner, Phys. Rev. A {\bf 40}, 4277 (1989)

 \bibitem{SZ_2016}
 J.Stolze and A.I.Zenchuk, Quant. Inf. Proc.  {\bf 15}, (2016) 3347

 \bibitem{BFZ_Arch2018}
G.A.Bochkin, E.B.Fel'dman, A.I.Zenchuk, Quant.Inf.Proc.  {\bf 17}, 218 (2018)

\bibitem{Z_JPA_2012}
A.I.Zenchuk, J. Phys. A: Math. Theor. {\bf 45} 115306 (2012) 


 \bibitem{FL}
 E. B. Fel'dman, S. Lacelle, Chem. Phys. Lett. {\bf 253}, 27
(1996)

\bibitem{FKZ_2010}
 E.B.Fel'dman, E.I.Kuznetsova and A.I.Zenchuk,
Phys. Rev. A {\bf 82}, 022332 (2010) 

\bibitem{LH}
Liu, L.L., Hwang, T., Controlled remote state preparation protocols via AKLT states,
Quantum Inf.Process. {\bf 13}, 1639-1650 (2014)
 
\bibitem{BZ_2016}
G. Bochkin and  A. Zenchuk, Extension of the remotely creatable  region via the local unitary
transformation on the receiver side, Quntum Information and Computation, {\bf 16}  (2016) 1349-1364

 \bibitem{FKZ_2016}
E.B.Fel'dman ,  E. I. Kuznetsova,  A.I.Zenchuk, 
Quantum Inf. Proc.  {\bf 15}  (2016) 2521

 \bibitem{BMGP}
J. Baum, M. Munowitz, A. N. Garroway, and A. Pines, J. Chem. Phys.
{\bf 83}, 2015 (1985).

\bibitem{DMF}
S. I. Doronin, I. I. Maksimov, and E. B. Fel'dman, J. Exp. Theor. Phys.
{\bf 91}, 597 (2000).

 \bibitem{DFZ_2011}
 S. I. Doronin,  E. B. Fel'dman, and A. I. Zenchuk, J.Chem.Phys. {\bf 134}, 034102 (2011)
 
\bibitem{KS1}
  H.G. Krojanski and D. Suter, Phys. Rev. Lett. {\bf 93}, 090501 (2004).

  \bibitem{KS2}
  H.G. Krojanski and D. Suter, Phys. Rev. Lett. {\bf 97}, 150503 (2006).
 
 \bibitem{AS}
 G. A. Alvarez and D. Suter, Phys. Rev. Lett. 104, 230403 (2010).
 
 \bibitem{CCGR}
 H.J. Cho, P. Cappellaro, D. G. Cory, and C. Ramanathan, Phys. Rev. B 74,
224434 (2006).

\bibitem{BFVV}
 G.A.Bochkin, E.B.Fel'dman, S.G.Vasil'ev, V.I.Volkov, Chem.Phys.Lett.
{\bf 680}, 56 (2017).

\bibitem{JW}
 P.Jordan, E.Wigner, Z. Phys. 47, 631 (1928)
 
 \bibitem{CG}
 H.B.Cruz, L.L.Goncalves, J. Phys. C: Solid State Phys. {\bf 14}, 2785 (1981)

\bibitem{Wootters}
 W.K. Wootters,
Phys. Rev. Lett. {\bf   80},
2245 (1998)

\bibitem{HW}
S.Hill and W.K.Wootters, Phys. Rev. Lett. {\bf    78}, 5022 (1997)

\bibitem{HV}
L.Henderson and V.Vedral J.Phys.A:Math.Gen. {\bf 34}, 6899 (2001)

\bibitem{OZ}
H.Ollivier and W.H.Zurek, Phys.Rev.Lett. {\bf 88}, 017901 (2001) 

\bibitem{Z}
W. H. Zurek, Rev. Mod. Phys. {\bf 75}, 715 (2003)



\bibitem{H}
Henry L. Haselgrove, Phys. Rev. A 72, 062326 (2005)

\bibitem{BOWB}
C. Allen Bishop, Yong-Cheng Ou, Zhao-Ming Wang, Mark S. Byrd, Phys.Rev. A {\bf 81}, 042313
(2010)

\bibitem{BZ_2016a}
G. Bochkin and  A. Zenchuk, arXiv:1612.02637v1

\bibitem{NJ}
G.M.Nikolopoulos and I.Jex, eds., {\it Quantum State Transfer and Network Engineering}, Series in
Quantum Science and Technology, Springer, Berlin Heidelberg (2014)

 \bibitem{SAOZ}
 J.Stolze, G. A. \'Alvarez,
O. Osenda, and  A. Zwick in
{\it Quantum State Transfer and Network Engineering.
Quantum Science and Technology},
ed. by  G.M.Nikolopoulos and I.Jex, Springer Berlin Heidelberg, Berlin, p.149  (2014)

\bibitem{ZZHE}
M.Zukowski, A.Zeilinger, M.A.Horne, and A.K.Ekert, Phys. Rev.
Lett. {\bf 71}, 4287 (1993)

\bibitem{BPMEWZ}
D.Bouwmeester, J.-W. Pan, K.Mattle, M.Eibl, H.Weinfurter, and  A. Zeilinger, 
Nature {\bf 390}, 575 (1997)

\bibitem{BBMHP}
D. Boschi,  S. Branca,  F. De Martini, L. Hardy,  and S. Popescu,
Phys. Rev. Lett. {\bf 80}, 1121 (1998)

\bibitem{DLMRKBPVZBW}
B.Dakic, Ya.O.Lipp, X.Ma, M.Ringbauer, S.Kropatschek,
S.Barz, T.Paterek, V.Vedral, A.Zeilinger, C.Brukner, and P.Walther, 
Nat. Phys. {\bf   8}, 666 (2012). 

\bibitem{DFZ_2011}
 S. I. Doronin,  E. B. Fel'dman, and A. I. Zenchuk, J.Chem.Phys. {\bf 134}, 034102 (2011)

\bibitem{BZ_arxiv2015}
G.A. Bochkin and A.I.Zenchuk, arXiv:1511.03507

\bibitem{DZ_arxiv2015}
S.I. Doronin and A.I. Zenchuk,  arXiv:1511.04331 

\bibitem{FKZ_arxiv2015}
E.B. Fel'dman, E.I. Kuznetsova, and A.I. Zenchuk, arXiv:1507.07738

\bibitem{ZASO22}
A.Zwick, G.A.Alvarez, J.Stolze, O.Osenda,
 Quant. Inf. Comput. {\bf 15},(2015) 582

\bibitem{ZAS}
 A. Zwick, G. A. Alvarez, J. Stolze, O. Osenda, Phys. Rev. A {\bf 85}, (2012) 012318

\bibitem{ZLZDLL}
J.Zhang, G. L. Long,  W. Zhang,  Zh. Deng,  W. Liu, and Zh. Lu, Phys.Rev.A {\bf 72}, 012331 (2005)

\bibitem{CRMF}
G. De Chiara, D. Rossini, S. Montangero, R. Fazio, Phys. Rev. A {\bf   72}, 012323 (2005)

\bibitem{ZASO}
 A. Zwick, G.A. \'Alvarez,
J. Stolze, O. Osenda, Phys. Rev. A {\bf   84}, 022311 (2011)
 
 \bibitem{ZASO2}
 A. Zwick, G.A. \'Alvarez,
J. Stolze, O. Osenda, Phys. Rev. A {\bf   85}, 012318 (2012)

\bibitem{ZASO3}
A. Zwick, G.A. \'Alvarez, J. Stolze, O Osenda,
Quant. Inf. Comput. {\bf 15}, 582 (2015)

\bibitem{P}
A.Peres, Phys. Rev. Lett. {\bf    77}, 1413 (1996)

\bibitem{AFOV}
L.Amico, R.Fazio, A.Osterloh and V.Ventral, Rev. Mod. Phys. {\bf    80}, 517 
(2008)

\bibitem{HHHH}
R.Horodecki,
P.Horodecki, M.Horodecki and K.Horodecki, Rev. Mod. Phys. {\bf    81}, 865  (2009)

\bibitem{DFZ}
S.I.Doronin, E.B.Fel'dman, and A.I.Zenchuk,
Phys. Rev. A {\bf 79}, 042310 (2009) 

\bibitem{DZ}
 S.I.Doronin, A.I.Zenchuk, Phys. Rev. A {\bf 81}, 022321 (2010)  	

\bibitem{LS}
P. Lorenz,  J. Stolze,
Phys. Rev. A {\bf 90}, 044301 (2014)

\bibitem{BBVB}
L.Banchi,  A. Bayat,  P. Verrucchi, and S.Bose, 
Phys.Rev.Let. {\bf 106}, 140501 (2011)

\bibitem{BZ}
 B. Chen, and  Zh. Song,
Sci. China-Phys., Mech. Astron {\bf 53}, 1266 (2010)

\bibitem{Banchi}
L. Banchi,
Eur. Phys. J. Plus {\bf 128}, 137 (2013) 

\bibitem{YGQ}
W. Qin, J. L. Li, G. L. Long,
Chin. Phys. B {\bf 24}, 040305 (2015) 

\bibitem{YGQ2}
Zh. Yang, M. Gao, W. Qin,
arXiv:1503.06274

\bibitem{QWZ}
W. Qin,  Ch. Wang,  and X. Zhang, Phys.Rev.A {\bf 91}, 042303 (2015)

\bibitem{Z_2012}
A.I.Zenchuk,
J. Phys. A: Math. Theor. {\bf   45} (2012) 115306

\bibitem{KF}
 E.I.Kuznetsova and E.B.Fel'dman, J.Exp.Theor.Phys. {\bf   102}, 882 (2006)

\bibitem{KZ_2008}
 E.I.Kuznetsova and A.I.Zenchuk, Phys.Lett.A {\bf   372},  pp.6134-6140 (2008)

\bibitem{FZ}
E.B.Fel'dman and A.I.Zenchuk,
Phys. Lett. A {\bf   373} (2009) 1719

\bibitem{BBCJPW}
C.H.Bennett, G.Brassard, C.Cr\'epeau, R.Jozsa, A.Peres, and W.K.Wootters,
Phys. Rev. Lett. {\bf   70}, 1895 (1993)

\bibitem{YS1}
B.Yurke, D.Stoler, Phys. Rev. A {\bf 46}, 2229 (1992)

\bibitem{YS2}
B.Yurke, D.Stoler, Phys. Rev. Lett. {\bf 68}, 1251 (1992)

\bibitem{BDSSBW}
C.H.Bennett, D.P.DiVincenzo, P.W.Shor, J.A.Smolin, B.M.Terhal, and W.K.Wootters,
Phys.Rev.Lett. {\bf   87}, 077902 (2001);
 Erratum,
 C.H.Bennett, D.P.DiVincenzo, P.W.Shor, J.A.Smolin, B.M.Terhal, and W.K.Wootters, 
 Phys. Rev. Lett. {bf   88}, 099902(E) (2002)

\bibitem{BHLSW}
C.H.Bennett, P.Hayden,
D.W.Leung, P.W.Shor, and A.Winter, 
IEEE Transetction on Information Theory {\bf   51}, 56 (2005)  

\bibitem{G}
G.L.Giorgi, Phys. Rev. A {\bf   88}, 022315 (2013) 

\bibitem{Z0}
W. H. Zurek, Ann. Phys.(Leipzig), {\bf 9}, 855 (2000)

\bibitem{DSC}
Datta, A., Shaji, A., Caves, C.M.,
Phys. Rev. Lett. {\bf 100}, 050502 (2008)

\bibitem{LBAW}
 Lanyon, B.P., Barbieri, M., Almeida, M.P., White, A.G.,
 Phys. Rev. Lett. {\bf 101}, 200501 (2008)

\bibitem{NLLZ}
W.J.Nie, Yu.H.Lan, Yo.Li, and Sh.Ya.Zhu, 
Sci.China-Phys., Mech. Astron {\bf 57}, 2276 (2014)

\bibitem{ZC}
P. Zhang, B. You, and L.-X. Cen,
Chin. Sci. Bull., {\bf 59}, 3841 (2014)

\bibitem{SXSZDWHCKW}
J.X. Sci, W. Xu, G. Sun et al,  Chin. Sci. Bull. {\bf 59}, 2547 (2014)

 \bibitem{RDL}
 S. Rodriques, N. Datta, and P. J. Love,
 Phys. Rev. A {\bf 90}, 012340 (2014) 